# Observation of intensity squeezing in resonance fluorescence from a solid-state device

Hui Wang, Jian Qin, Si Chen, Ming-Cheng Chen, Xiang You, Xing Ding, Y.-H. Huo, Ying Yu, C. Schneider, Sven Höfling, Marlan Scully, Chao-Yang Lu, Jian-Wei Pan

[1] Hefei National Laboratory for Physical Sciences at Microscale and Department of Modern Physics, University of Science and Technology of China, Hefei 230026, China

[2] Shanghai branch, CAS Centre for Excellence and Synergetic Innovation Centre in Quantum Information and Quantum Physics, University of Science and Technology of China, Shanghai 201315, China

[3] Technische Physik, Physikalisches Institüt and Wilhelm Conrad Röntgen-Center for Complex Material Systems, Universitat Würzburg, Am Hubland, D-97074 Würzburg, Germany

[4] SUPA, School of Physics and Astronomy, University of St. Andrews, St. Andrews KY16 9SS, United Kingdom

[5] Institute for Quantum Science and Engineering, Texas A&M University, College Station, Texas 77843, USA

[6] Department of Physics, Baylor University, Waco, Texas 76798, USA

[7] Department of Mechanical and Aerospace Engineering, Princeton University, Princeton, New Jersey 08544, USA

## Abstract

**Intensity squeezing[1]—i.e., photon number fluctuations below the shot noise limit—is a fundamental aspect of quantum optics and has wide applications in quantum metrology[2-4]. It was predicted in 1979 that the intensity squeezing could be observed in resonance fluorescence from a two-level quantum system[5]. Yet, its experimental observation in solid states was hindered by inefficiencies in generating, collecting and detecting resonance fluorescence. Here, we report the intensity squeezing in a single-mode fibre-coupled resonance fluorescence single-photon source based on a quantum dot-micropillar system. We detect pulsed single-photon streams with 22.6% system efficiency, which show subshot-noise intensity fluctuation with an intensity squeezing of 0.59 dB. We estimate a corrected squeezing of 3.29 dB at the first lens. The observed intensity squeezing provides the last piece of the fundamental picture of resonance fluorescence; which can be used as a new standard for optical radiation and in scalable quantum metrology with indistinguishable single photons.**

A fundamental tool in optical quantum information science is the single-photon source[6-7], which ideally should emit one, and only one, indistinguishable light quanta on demand. The absence of two- or more-photon components gives rise to antibunching—a highly nonclassical behaviour—which is relevant for quantum cryptography applications[8]. The indistinguishability between independent single photons, which can be measured by two-photon Hong-Ou-Mandel quantum interference[9], lies at the heart of quantum teleportation[10], quantum networks[11], and linear optical quantum computing[12,13].

In principle, an ideal single-photon device is also useful for quantum metrology. For example, there is a growing interest in redefining the international standard base unit of luminous intensity, candela, by counting the number of fundamental light quanta, photons[14]. Compared to conventional light sources, such as lasers or light-emitting diodes, an ideal single-photon source has the unique quantum advantage that its intensity uncertainty can be greatly suppressed, thus offer a way to beat the fundamental shot noise limit. Such a feature would find applications in various fields, such as imaging or measurement in the ultralow power regime of light-sensitive biological samples[15,16].

While the high single-photon purity, indistinguishability, and extraction efficiency has been demonstrated[17-19], the simultaneous observation of intensity squeezing in a single-photon source remains challenging. This is due to the inevitable photon loss from the generation, transmission, collection, and detection in experiments; such that, in each time slot, there was much more often vacuum rather than one photon detected. While the low overall efficiency only affected the data accumulation time in the measurements of the single-photon purity and indistinguishability, the intensity squeezing, which heavily relies on the overall efficiency; can be diminished by the low efficiency and washed out in the presence of fluctuation of the experimental parameters. Therefore, there were rarely observations of intensity squeezing in single-photon sources. The first observation, with a very small squeezing of ~0.00183±0.00038 dB, dated back to 1983 from a single atom[1]. Recently, a single molecule embedded inside a metallodielectric antenna was designed for a strong enhancement of photon collection efficiency and showed a squeezing of 2.2 dB[20]. However, these single photons were not in a single mode, and due to the incoherent excitation, they were distinguishable. A related work reported intensity squeezing with confined polaritons in semiconductor micropillars[21].

Recently, semiconductor quantum dots embedded in microcavities emerged as a scalable solid-state platform for quantum information technologies[22-23]. Pulsed and resonant excitation of the single quantum dots was used to efficiently generate single photons with near-unity purity and indistinguishability[24]. Polarized microcavities were deterministically coupled to the emitter and efficiently funnelled the single photons into a single spatial mode[25]. In all physical systems, the long sought-after goal of observing intensity squeezing in pulsed resonance fluorescence remained a challenge. Here, we report the first single-mode fibre-coupled semiconductor source of indistinguishable single photons with subshot-noise intensity fluctuation.

The reduction of the intensity fluctuation in a single-photon source compared to the shot noise can be expressed as[20]: $N_{SPS}/N_{SN}=\sqrt{1-\rho T}$, where $\rho$ is the internal efficiency of the quantum emitter (including its quantum efficiency and excited-state preparation efficiency), and $T$ denotes the total external efficiency (including extraction efficiency and the overall transmission efficiency in the experiment setup). Due to the quantum uncertainty principle, when the uncertainty of the intensity is smaller than the shot noise, the uncertainty of its conjugate variable—phase—is accordingly increased.

In our experiment, we use a self-assembled semiconductor InAs/GaAs quantum dot coupled in a 2-μm diameter micropillar cavity. The sample is cooled at 4 K where the emitter is resonant with the cavity, so that both the efficiency and indistinguishability are improved due to the Purcell effect. A confocal microscope is utilized to excite the emitter and collect resonance fluorescence single photons. A cross-polarization set-up suppresses the laser leakage with an extinction ratio of ~$10^7$:1.

First, we study the resonance fluorescence single photons under continuous-wave (CW) laser excitation. Due to the high photon flux and the limited recovery time (~5 ns) of the superconducting nanowire single-photon detectors used, we attenuate the single-photon stream by 1000 times. The corrected single-photon count rate as a function of laser power is shown in Fig. 1(a), from which we extract a saturation power of $P_{sat} = 4.9$ nW and a saturated photon flux of $I_\infty = (1.87 \pm 0.03) \times 10^9 / s$. This is so far the brightest single-photon source reported in any physical system.

A CW laser can prepare the emitter in its excited state with a maximal probability of 50% in the limit of high laser power[26]. The theoretically predicted single-photon flux is $1/2T_1$, where $T_1$ is the excited state lifetime. Fig. 1(b) shows a time-resolved resonance fluorescence measurement which gives $T_1 = 58.60 \pm 0.02$ ps. Thus the overall single-photon efficiency is $\frac{I_\infty}{1/2T_1} = 0.219 \pm 0.004$. Compared to the average lifetime of many quantum dots in the slab, $T_{slab} = 1.08$ ns, we estimate the Purcell factor to be ~18.4.

Next, we resonantly pump the quantum dot by a pulsed laser with a bandwidth of ~50 GHz to match the cavity mode. The detected single-photon counts show a Rabi oscillation as a function of excitation power, as plotted in Fig. 1(c). At π pulse, with a repetition rate of 76 MHz, we finally detect 17.2 million single photons per second on a superconducting nanowire single-photon detector with an efficiency of ~86%. The corresponding overall single-photon efficiency is 22.6%, in a good agreement with the value (21.9%) extracted from CW excitation.

The single photon nature of generated resonance fluorescence is demonstrated by a Hanbury Brown and Twiss measurement that shows $g_2(0) = 0.025(1)$ at a π pulse. Fig. 1(d) is the measured high-resolution spectrum of the resonance fluorescence single photons, showing a full width at half maximum of $2.74 \pm 0.02$ GHz. By fast Fourier transformation, the fitted coherent time is $T_2 = 108.8 \pm 0.9$ ps. This allows us to estimate that $T_2/2T_1 = 0.928 \pm 0.008$, indicate that the pulsed single photons are 92.8% Fourier transform limit. This is in a good agreement with the measured photon indistinguishability via Hong-Ou-Mandel interferences that show a corrected visibility of 0.935(1) for two single photons separated by ~10 μs.

We now examine the intensity squeezing of resonance fluorescence. The single photons are directly detected by a superconducting nanowire single-photon detector, and the arrival time is recorded by a time-to-digit converter. Figure 2(a) is the real-time monitoring of resonance fluorescence single photon counts at π pulse with a time bin of 1.0 μs. The average count per time bin is 17.5/μs. The corresponding histogram is plotted in Fig. 2b. The directly observed standard deviation of photon counts is $N_{SPS} = 3.65$, which is significantly below the shot noise limit of $N_{SN} = \sqrt{17.48} = 4.18$. Therefore, the directly measured intensity squeezing is $N_{SPS}/N_{SN} = 87.32\%$ (0.59 dB) at the π pulse. This value is very close to, but slightly smaller than the theoretically predicted value of 87.75%. This is due to the imperfect recovery time of the detector (see supplementary information). As a comparison experiment, we replace the single-photon source by a CW laser attenuated to an average count rate of ~17.2 million per second and send it into the same detector. The observed standard deviation is 99.98% of the short noise limit. This indicates that the observed subshot-noise intensity fluctuation indeed comes from the single photons.

In Fig. 2(b), the occurrences at different single-photon counts are plotted (deep blue) and fitted by a binominal function (see Supplementary Information), which is reduced by 12.68% compared to the shot noise limit (green). Considering the optical transmission loss, cross-polarization filtering loss, coupling loss into the single-mode fiber, and the detection efficiency at the detector, the intensity squeezing out of the first lens should be 3.29 dB (light blue). Fig. 2(c) shows the amount of intensity squeezing as a function of the excitation laser amplitude. As expected, by tuning the laser power gradually to π pulse, the excited-state population—proportional to the single-photon generation efficiency—grows, which increases the amount of intensity squeezing.

It is clear that the intensity squeezing critically relies on the overall efficiency. We analyse the photon loss budget in detail. First, we estimate the external efficiency, which accounts for all photon loss from the first lens to the single-photon detector:

collection efficiency by the first objective, polarization filtering, single-mode fibre coupling, and detection. To estimate the single-mode fibre coupling efficiency, we performed a separate experiment using multimode fibres (with diameters of 100 μm, 200 μm, which gives the same counts, indicating the efficiency is saturated and equal to unity). The ratio of single-photon count using a single-mode fibre is measured to be 74% of that using multimode fibres, which determines the single-mode coupling efficiency. The overall transmission efficiency of the optical path (including the first lens, optical window, polarized beam splitter, and two wave plates) is measured to be 83%. The first lens collection efficiency is estimated to be 78%. The superconducting nanowire single-photon detection efficiency is 86%. The micropillar cavity shows a small mode splitting due to a slight ellipticity of its cross section. In this case, the intensity of one eigenmode will be higher than that of another one[32]. By measuring the single photon counts in two modes respectively, the determined polarization filtering efficiency is 55%.

The internal efficiency of the quantum dot-micropillar system includes the quantum efficiency of the emitter (QE), excited-state preparation efficiency (ESPE), and the photon extraction efficiency (PEE) out of the micropillar. The QE and ESPE can be affected by power-induced[27] and phonon-induced damping[28] in quantum dots. The PEE accounts for the losses due to the side leakage and the part not coupled to the fundamental mode of the microcavity, which can be estimated by[29]

$$\eta_C = \frac{F_p}{F_p + 1} \frac{Q}{Q_0},$$

where $F_p$ is the Purcell factor, $Q$ is the quality factor, and $Q_0$ is the quality factor of planar microcavity. In this experiment, $F_p=18.4$, $Q=6800$, $Q_0=7600$, so the photon extraction efficiency by the microcavity is ~85%. This result is consistent with the numerically simulated number of 87% using finite difference time domain. The product of QE and ESPE is estimated to be 92%.

In summary, we have directly observed the first intensity squeezing in pulsed resonance fluorescence, in a single-mode fibre-coupled highly-indistinguishable solid-state single-photon source. The uncorrected squeezing at the output of the single-mode fibre is 0.59 dB, and the corrected intensity squeezing at the first lens is 3.29 dB. From a fundamental perspective, our work fills the a long sought-after missing element in textbook quantum optical phenomena[26] using quantum dots. This is addition to the previously reported antibunching[30], two-photon interference[31], weak[32] and strong[33,34] coupling, Rabi oscillation[35], Autler-Townes splitting[36], coherent population trapping[37], Mollow triplet[38,39], and quadrature squeezing[40]. For practical applications, our work combines for the first time high levels of single-photon efficiency, purity and indistinguishability together with intensity squeezing in a semiconductor chip. The intensity squeezing can be further improved by driving the quantum dot with dichromatic laser pulses[41] and/or coupling the quantum dot to polarized microcavities[19]. We note that such an intensity squeezed single-photon source with subshot noise fluctuations has natural application to benchmarking single-photon detector efficiency[42], redefining the standard base unit of luminous intensity at ultralow power level[43], and optical spectroscopy of light-sensitive biological samples[15,16].

**References**

1. Short, R., & Mandel, L. Observation of sub-Poissonian photon statistics. *Phys. Rev. Lett.* **51**, 384 (1983).
2. Boto, Agedi N., *et al*. Quantum interferometric optical lithography: exploiting entanglement to beat the diffraction limit. *Physical Review Letters* **85**.13 (2000): 2733.
3. Giovannetti, Vittorio, Seth Lloyd, and Lorenzo Maccone. Quantum metrology. *Physical Review Letters* **96**.1 (2006): 010401.
4. Giovannetti, V., Lloyd, S, and Maccone, L., Advances in quantum metrology, *Nature Photon.* **5**, 222-229 (2011).
5. Mandel, L., Sub-Poissonian photon statistics in resonance fluorescence. *Opt. Lett.* **4**, 205-207 (1979).
6. Lounis, B. & Orrit, M. Single-photon sources. *Rep. Prog. Phys*. **68**,1129–1179 (2005).
7. Shields, A. J. Semiconductor quantum light sources, *Nature Photon.* **1**, 215-223 (2007).
8. Bennett, C. H., and Brassard, G., Quantum cryptography: Public key distribution and coin tossing, In *Proceedings of IEEE International Conference on Computers, Systems and Signal Processing*, 175-179 (1984).
9. Hong, C. K., Ou, Z. Y., and Mandel, L. Measurement of subpicosecond time intervals between two photons by interference. *Phys. Rev. Lett.* **59**, 2044-2046 (1987).

10. Bennett, C. H., Brassard, G., Crepeau, C., Jozsa, R., Peres, A., and Wootters, W. K., Teleporting an unknown quantum state via dual classical and Einstein-Podolsky-Rosen channels, *Phys. Rev. Lett.* **70**, 1895 (1993).
11. Kimble, H. J., The quantum internet. *Nature* **453**, 1023-1030 (2008).
12. Kok, P., Munro, W. J., Nemoto, K., Ralph, T. C., Dowling, J. P. & Milburn, G. J. Linear optical quantum computing with photonic qubits. *Rev. Mod. Phys.* **79**, 135-174 (2007).
13. Pan, J. W., Chen, Z.-B., Lu, C. Y., Weinfurter, H., Zeilinger, A. & Żukowski, M. Multiphoton entanglement and interferometry. *Rev. Mod. Phys.* **84**, 777-838 (2012).
14. Cheung, J. Y. *et al*. The quantum candela: a re-definition of the standard units for optical radiation. *J. Mod. Opt.* **54**, 373-396 (2007).
15. Taylor, M., Janousek, J., Daria, V. *et al*. Biological measurement beyond the quantum limit. *Nature Photon* **7**, 229–233 (2013).
16. Tenne, R., Rossman, U., Rephael, B. *et al*. Super-resolution enhancement by quantum image scanning microscopy. *Nature Photon* **13**, 116–122 (2019).
17. Ding, X. *et al.* On-Demand single photons with high extraction efficiency and near-unity indistinguishability from a resonantly driven quantum dot in a micropillar. *Phys. Rev. Lett.* **116**, 020401 (2016).
18. Somaschi, N. *et al*. Near-optimal single-photon sources in the solid state. *Nature Photon.* **10**, 340-345 (2016).
19. Wang, H. *et al.*, Towards optimal single-photon sources from polarized elliptical microcavities. *Nature Photon.* **13**, 770-775 (2019).
20. Chu, X. L., Gotzinger, S. & Sandoghdar, V. A single molecule as a high-fidelity photon gun for producing intensity-squeezed light. *Nature Photon.* **11**, 58-62 (2017)
21. Boulier, T., Bamba, M., Amo, A. et al. Polariton-generated intensity squeezing in semiconductor micropillars. *Nat Commun.* **5**, 3260 (2014)
22. Senellart, P., Solomon, G. & White, A. High-performance semiconductor quantum-dot single-photon sources. *Nature Nanotech.* **12**, 1026-1039 (2017).
23. Wang, H. *et al*. Boson sampling with 20 input photons and a 60-mode interferometer in a $10^{14}$-dimensional Hilbert space. *Phys. Rev. Lett.* **123**, 250503 (2019).
24. He, Y.-M. *et al.* On-demand semiconductor single-photon source with near-unity indistinguishability. *Nature Nanotech.* **8**, 213-217 (2013).
25. Wang, H. *et al.*, Towards optimal single-photon sources from polarized elliptical microcavities. *Nature Photon.* **13**, 770-775 (2019).
26. Scully M. O. & Zubairy M. S. Quantum optics. 1999.
27. Mogilevtsev, D. *et al*. Driving-dependent damping of Rabi oscillations in two-level semiconductor systems. *Phys. Rev. Lett.* **100**, 017401 (2008).
28. Ramsay, A. J. *et al*. Damping of exciton rabi rotations by acoustic phonons in optically excited InGaAs/GaAs quantum dots. *Phys. Rev. Lett.* **104**, 017402 (2010).
29. Barnes, W. L. *et al*. Solid-state single photon sources: light collection strategies. *Eur. Phys. J.*

*D* **18**, 197–210 (2002).

30. Michler, P. *et al*. A quantum dot single-photon turnstile device. *Science* **290**, 2282–2285 (2000).
31. Santori, C., Fattal, D., Vučković, J. *et al*. Indistinguishable photons from a single-photon device. *Nature* **419**, 594–597 (2002)
32. Gérard, J. M. *et al*. Enhanced spontaneous emission by quantum boxes in a monolithic optical microcavity. *Physical Review Letters* **81**.5 (1998): 1110.
33. Yoshie, T., Scherer, A., Hendrickson, J. *et al*. Vacuum Rabi splitting with a single quantum dot in a photonic crystal nanocavity. *Nature* **432**, 200–203 (2004).
34. Reithmaier, J., Sęk, G., Löffler, A. *et al*. Strong coupling in a single quantum dot–semiconductor microcavity system. *Nature* **432**, 197–200 (2004).
35. Stievater, T. H., *et al*. Rabi oscillations of excitons in single quantum dots. *Physical Review Letters* **87**.13 (2001): 133603.
36. Xu, X. *et al*. Coherent optical spectroscopy of a strongly driven quantum dot. *Science* **317**.5840 (2007): 929-932.
37. Xu, X., Sun, B., Berman, P. *et al*. Coherent population trapping of an electron spin in a single negatively charged quantum dot. *Nature Phys* **4**, 692–695 (2008).
38. Nick Vamivakas, A., Zhao, Y., Lu, C. *et al*. Spin-resolved quantum-dot resonance fluorescence. *Nature Phys* **5**, 198–202 (2009).
39. Flagg, E., Muller, A., Robertson, J. *et al*. Resonantly driven coherent oscillations in a solid-state quantum emitter. *Nature Phys* **5**, 203–207 (2009).
40. Schulte, C., Hansom, J., Jones, A. *et al*. Quadrature squeezed photons from a two-level system. *Nature* **525**, 222–225 (2015).
41. He, Y.-M. *et al.*, Coherently driving a quantum two-level system with dichromatic laser pulses. *Nature Phys.* **15**, 941-946 (2009).
42. Hadfield, R. Single-photon detectors for optical quantum information applications. *Nature Photon* **3**, 696–705 (2009).
43. Polyakov, S. V. & Migdall, A. L. Quantum radiometry. *J. Mod. Opt.* **56**, 1045–1052 (2009).

**Figure captions:**

Figure 1: (a) Single-photon counts with CW laser resonant excitation. The saturation count is 1.87 billion per second. The data is fitted by the Eqn. (2). (b) Time-resolved measurement of the resonance fluorescence by a superconducting nanowire single photon detector with a time resolution of ~20 ps. The fitted lifetime of quantum dot is $T_1 = 58.60 \pm 0.02$ ps. Comparing to the quantum dot lifetime in the slab (~1.08 ns), the Purcell factor is 18.4. (c) Rabi oscillation under pulsed resonant excitation. At π pulse, 17.2 million pure single photons are detected per second, by a superconducting nanowire single-photon detector with an efficiency of 0.86. (d) High-resolution spectrum of resonance fluorescence measured by a Fabry-Perot cavity with a frequency

resolution of 220 MHz and a free spectral range of 37.4 GHz. The data is fitted by a Voigt function (red line), and the linewidth is $2.74 \pm 0.02$ GHz.

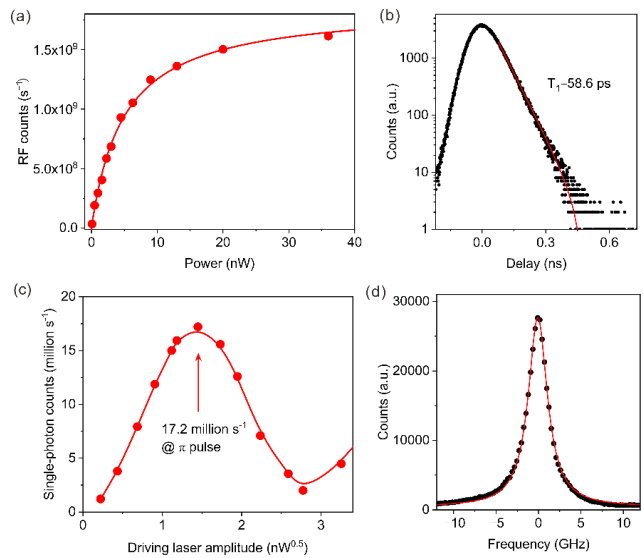

Figure 2: (a) Real-time monitoring of resonance fluorescence single photon counts at π pulse with a time bin of 1.0 μs. The average count per time bin is 17.5/μs. (b) The corresponding histogram where the frequency of observed counts (i.e., the intensity fluctuation, shown as deep blue dot) is fitted by a binomial function. Comparing to the shot-noise-limited source with the same intensity (green line), the quantum dot single-photon source shows a 12.68% reduction of histogram linewidth, that is, intensity fluctuation noise. Blue line displays the corrected intensity squeezing at the first lens (see main text). (c) Intensity squeezing parameter as a function of the excitation laser amplitude. At π pulse, the intensity squeezing reaches a minimal value of 87.32%. Black dotted line is the normalized shot-noise limit.

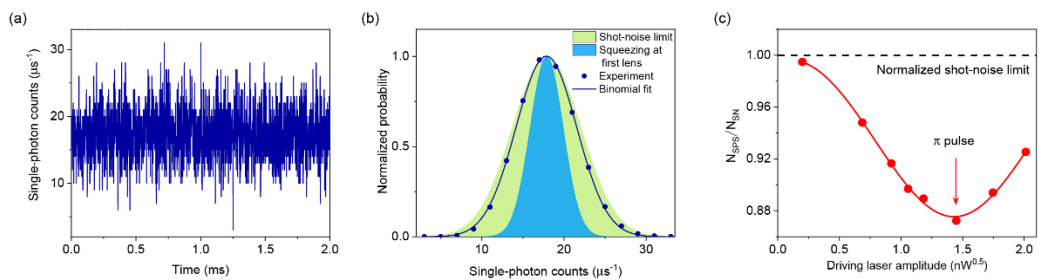